\documentstyle[11pt,aaspp4]{article}
\tighten
\eqsecnum

\begin{document}

\font\twelvei = cmmi10 scaled\magstep1 
       \font\teni = cmmi10 \font\seveni = cmmi7
\font\mbf = cmmib10 scaled\magstep1
       \font\mbfs = cmmib10 \font\mbfss = cmmib10 scaled 833
\font\msybf = cmbsy10 scaled\magstep1
       \font\msybfs = cmbsy10 \font\msybfss = cmbsy10 scaled 833
\textfont1 = \twelvei
       \scriptfont1 = \twelvei \scriptscriptfont1 = \teni
       \def\mit{\fam1 }
\textfont9 = \mbf
       \scriptfont9 = \mbfs \scriptscriptfont9 = \mbfss
       \def\bmit{\fam9 }
\textfont10 = \msybf
       \scriptfont10 = \msybfs \scriptscriptfont10 = \msybfss
       \def\bmsy{\fam10 }

\def\etal{{\it et al.~}}
\def\eg{{\it e.g.,~}}
\def\ie{{\it i.e.,~}}
\def\lsim{\raise0.3ex\hbox{$<$}\kern-0.75em{\lower0.65ex\hbox{$\sim$}}} 
\def\gsim{\raise0.3ex\hbox{$>$}\kern-0.75em{\lower0.65ex\hbox{$\sim$}}} 
\def\cf{{\it cf.~}}
\def\cc{CCs}
\def\tt{ $t$ }
\def\coll{ $\tau_{coll}$}
\def\rad{ $\tau_{rad}$}
\def\tcr{ $\tau_{cr}$}
\def\rmm{ $\rho_{max}$}
\def\pm{ $\p_{max}$}
\def\Pac{{Paczy\'{n}ski\/}}
\def\ltsima{$\; \buildrel < \over \sim \;$}
\def\simlt{\lower.5ex\hbox{\ltsima}}
\def\gtsima{$\; \buildrel > \over \sim \;$}
\def\simgt{\lower.5ex\hbox{\gtsima}}

\title{Hydrodynamics of Cloud Collisions in 2D: \\The Fate of Clouds in a Multi-phase Medium\altaffilmark{10}}

\author{Francesco Miniati \altaffilmark{1,2,6},
        T. W. Jones\altaffilmark{1,7},
        Andrea Ferrara \altaffilmark{3,8},
and     Dongsu Ryu \altaffilmark{4,5,9}}

\altaffiltext{1}{Department of Astronomy, University of Minnesota,
    Minneapolis, MN 55455}
\altaffiltext{2}{Dipartimento di Astronomia, Universit\`a of Firenze,
    50125 Firenze, Italy}
\altaffiltext{3}{Osservatorio Astrofisico di Arcetri,
    50125 Firenze, Italy}
\altaffiltext{4}{Department of Astronomy \& Space Science, Chungnam National
    University, Daejeon 305-764, Korea}
\altaffiltext{5}{Department of Astronomy, University of Washington,
    Box 351580, Seattle, WA 98195-1580}
\altaffiltext{6}{e-mail: min@msi.umn.edu}
\altaffiltext{7}{e-mail: twj@astro.spa.umn.edu}
\altaffiltext{8}{e-mail: ferrara@arcetri.astro.it}
\altaffiltext{9}{e-mail: ryu@sirius.chungnam.ac.kr}
\altaffiltext{10}{Published in {\it The Astrophysical Journal}, December 10, 1997}

\begin{abstract}

We have studied head-on collisions between equal-mass, mildly supersonic HI clouds
(Mach number 1.5 with respect to the background medium)
through high resolution numerical simulations in two dimensions.
We explore the r\^ole of 
various factors, including the radiative cooling parameter, $\eta$ = \rad/\coll~
(\coll=$R_c/v_c$), evolutionary modifications on the cloud structure
and the symmetry of the problem.
Self-gravity is not included. Radiative losses are taken into account 
explicitly and not approximated with an isothermal adiabatic index 
$\gamma\approx1$, which, in fact, leads to very different results.
We assume a standard two-phase interstellar medium (ISM)
model where clouds are characterized by a temperature
$T_c = 74 K$ and number density
$n_c = 22 cm^{-3}$ and are in pressure equilibrium with the surrounding
warm intercloud medium (WIM), with a density
contrast $\chi=\rho_c/\rho_i=100$.
In particular, we study collisions for the adiabatic ($\eta \gg 1$) and radiative ($\eta =0.38$)
cases which
may correspond to small ($R_c\le 0.4pc$ for assumed WIM) or large ($R_c\sim 1.5pc$)
clouds, respectively. In addition to a standard case of 
identical ``non-evolved'' clouds, we also consider
the collision of identical clouds, ``evolved'' through independent
motion within the intercloud gas, over one crushing time before
collision.
This turns out to be about the mean collision time for
such clouds in the ISM.
The presence of bow shocks and ram pressure from material 
in the cloud wake, significantly alters these interactions with respect to the 
standard case.
In some cases, we removed the mirror symmetry from the problem by colliding initially
identical clouds
``evolved'' to different ages before impact.
In those cases the colliding clouds have different density and
velocity structures, so that they provide a first insight on the behavior of
more complex interactions.

In our adiabatic collisions the clouds are generally disrupted
and convert their gas into the warm 
phase of the ISM. Although the details depend on the initial conditions, 
the two colliding clouds are converted into a few low
density contrast ($\chi\sim 5$) clumps at the end of the simulations.

By contrast, for {\it symmetric} radiative cases we find that
the two clouds coalesce, and there are good 
chances for a new massive cloud to be formed. Almost all the 
initial kinetic energy of the two clouds is radiated away during
such collisions.
On the other hand, for both adiabatic and radiative collisions,
symmetry breaking leads to major differences.
Most importantly, asymmetric collisions have a much greater
tendency to disrupt the two clouds. Portions of individual clouds
may be sheared away, and instabilities along the interfaces between
the clouds and with the intercloud medium are enhanced. In addition, 
radiative cooling is less efficient in our asymmetric interactions, so that those
parts of the clouds that initially seem to merge are more likely to
re-expand and fade into the warm intercloud medium.
Since the majority of real cloud collisions should be asymmetric for
one reason or another, we conclude that most gasdynamical
diffuse cloud collisions will be disruptive, at least in the absence of
significant self-gravity or of a significant magnetic field.

\end{abstract}


\clearpage

\section{Introduction}

Interstellar cloud collisions (\cc~ hereafter) are important for the
dynamical evolution of galaxies. Indeed, \cc~ turn out to be relevant
in a large variety of important processes occurring
in the interstellar medium (ISM),
such as star formation, dissipation of kinetic energy and 
gas phase transitions. In addition, they
affect the gaseous structure and the energy budget of 
galaxies, along with the mass spectrum of diffuse ISM clouds and their evolution.

\cc~were first studied by Stone (1970a, 1970b), who
was mainly concerned with the 
evolution of interstellar clouds and especially of their structure.
In his pioneering
papers he found that, despite the very high compression undergone
by clouds during \cc, star formation is not enhanced.
Instead, he found that colliding clouds lose mass and that large scale perturbations and internal
motions (approximately radial) form within the clouds, which do not decay before 
the next collision. Thus, cloud
hydrostatic equilibrium is severely compromised, suggesting the idea
of clouds with smooth density distribution profiles
instead of (clouds as) sharply bounded objects (Stone 1970b).
Smith (1980) performed 1D numerical simulations and concluded that,
in a dusty ISM, low Mach number
($\le$ 5) collisions are the most likely to trigger star formation.
Further investigations have been carried out by Gilden (1984) and 
Lattanzio \etal (1985, LMPS hereafter) using 2D and 3D models, respectively. These authors 
studied the r\^ole of various parameters involved in the collisions, such  as
the relative size of the clouds, their masses and impact parameter.
They generally concluded that \cc~more often lead to disruption than to
coalescence or gravitational collapse of the clouds.
This again has been interpreted as evidence that the ISM should be
described as
a continuous distribution of gas streams and turbulent eddies (LMPS,
Hunter \etal, 1985). That contrasts with most dynamical models for the 
ISM that depict it more simply as pressure-confined neutral cold clouds (CNM) moving
through a warm diffuse inter-cloud
medium (WIM) (see \eg Norman and Ferrara 1996, 
Vazquez-Semadeni \etal 1995 and references therein). This point is so important that
further investigation of the final fate of the clouds is worthwhile. 
In fact, it turns out that \cc~are very frequent, with a rate of
about once every $10^6~-~10^7yr$ for individual clouds, leading to
a total of $\sim 1$ cloud collision every 100 $yr$
in the Galaxy (Stone 1970, Spitzer 1978, Hausman 1981).
Their outcome could profoundly affect
the mass interchange of the various phases that are believed to exist in the ISM.

\cc~ may also be responsible for the buildup of the
observed mass spectrum of diffuse clouds 
(Dickey \& Garwood 1989,
Solomon \& Rivolo 1989, Oort 1954,
Field \& Saslaw 1965, Field \& Hutchings 1968, Penston \etal 1969, Cowie
1980, Hausman 1982, Pumphrey \& Scalo 1983, Struck-Marcell \& Scalo 1984,
Fleck 1996, Mousumi \& Chanda 1996).
Early models suggested a hierarchical scenario in which small clouds, formed
in HII regions out of much larger clouds
through star formation, are supposed to
undergo repeated inelastic collisions and coalescence,
thus engendering new clouds of larger mass (Oort 1954, Field \& Saslaw 1965).
Yet, an accurate model should take into account the detailed outcome of CC~
processes. In their model Field \& Saslaw (1965) assume, for example,
that each CC~leads to coalescence, an uncertain assumption as discussed above
and by several authors including
Stone (1970a, 1970b), Hausman (1981), Gilden (1984)
and Klein, McKee \& Wood (1994, KMW hereafter). 

In addition, \cc~are relevant for the energy budget and gaseous structure 
of a galaxy.
Clouds are accelerated in HII regions (Field \& Saslaw
1965) and by supernovae. The turbulence pressure associated with their motion
determines the vertical structure of the cold neutral phase in the ISM
(McKee 1990, Ferrara 1993, Norman \& Ferrara 1996).
The bulk motions, on the other hand, are dissipated through
inelastic collisions (Spitzer 1978). 
It is also worthwhile to mention that the amount of energy dissipated during
\cc~depends on the dust content and on the metallicity of the ISM,
as discussed by Ricotti, Ferrara \& Miniati  (1997, RFM hereafter), who
determined the dependence of the 
elasticity of a collision on parameters such as relative cloud velocity,
size, metallicity and dust-to-gas ratio. Their paper gives a useful overview 
of the characteristics of collisions and represents a complementary study to 
the present one.

The present paper is devoted to a further investigation of the consequences
of \cc. In particular, we hope to determine the fate (i.e. survival, dispersal
in the WIM, coalescence or fragmentation) of clouds in a two-phase ISM with
canonical properties.  
In addition, we investigate the evolution of the cloud kinetic energy.
This could be retained by the clouds as thermal energy, radiated away 
during the collision, or transferred to the WIM.
Very recently some attention has been devoted to the hydrodynamical
details that characterize the flow structure in \cc~(KMW, Kimura \& Tosa 1996). 
KMW
have pointed out important instabilities
that develop in the collision process. In particular, they have shown that 
small amplitude surface perturbations may lead to the 
development of the ``bending mode instability'' (Vishniac 1994)
in the cloud shocks,
which, in turn, causes
the colliding material to form filamentary structures.
Our work extends the KMW study by
adding individual cloud evolution through the intercloud medium,
before the collision. 
This increases the realism of the simulations by
taking into account the r\^ole of cloud bow shocks and wakes and of 
nonuniform cloud structures. By colliding two clouds that have undergone
different amounts of prior evolution, we introduce a simple means to relax 
the symmetry imposed in most previous calculations.
This turns out to alter dramatically the result of the interactions
in ways that extend beyond the direct influence of the bending mode.
We still limit for now our 
attention to head-on 2D gasdynamical collisions between clouds of 
(initially) equal mass and Mach number.

The plan of this paper is as follows. In \S 2 we describe the physical problem
under investigation and discuss the assumptions made.
In \S 3 we describe the computational setup;
\S 4 presents the main results of this study; 
discussion is given in \S 5 whereas summary and conclusions are in \S 6.

\section{Description of the problem}
\label{descr}

It is known (\eg, Stone 1970a, 1970b, KMW) that 
collisions between identical homogeneous clouds are generally characterized
by three main evolutionary stages as shown in Fig. 1. These are
(1) a {\it compression phase} in which 
a shock wave forms in each cloud and propagates through it, converting cloud
kinetic energy into thermal energy. If radiative cooling is efficient, a fair
fraction of this thermal energy is radiated away. Otherwise, the pressure
becomes high enough to drive shocks outward from the center of the collision.
At the end of the compression phase
the gas is highly compressed into a thin (compared to the initial cloud size)
layer, whose thickness depends on the amount of radiative losses (Fig. \ref{fcat0}a
and top panel of Fig. \ref{fcct0}a). 
(2) A {\it re-expansion phase} begins when shocks generated during the
collision emerge into the inter-cloud medium.
During this phase a rarefaction wave
propagates backwards into the clouds, forming a central low pressure and low 
density region
(Fig. \ref{fcat0}b and 2). Also, a fast sheet of ejecta  emerges from the contact
discontinuity between the merged clouds. 
This structure, which appears like a ``jet'' in 2D collisions
with mirror symmetry between the clouds, is subject to
Kelvin-Helmholtz instabilities (KHIs) (Fig. \ref{fcct0}b).
Also, during the re-expansion phase, a reverse shock forms and interacts with 
the expanding cloud material.
As will be shown in \S \ref{resu},
the qualitative and quantitative details of this phase strongly depend
on the adiabaticity  and symmetry of the collision. 
(3) There follows a {\it collapse phase} when the expansion is halted by
the external surrounding medium. The pressure inside the clouds 
is by now much lower than in the external ambient medium, so that
the cloud material is driven back toward the merged cloud core.
As pointed out in the literature, 
during this phase Rayleigh-Taylor instabilities (RTIs) become very important,
especially in disrupting the cloud surface and ablating material away from it
(Fig. \ref{fcat0}c).
The details of the evolution of these properties are obviously closely related to the
previous, re-expansion phase
and thus depend as well on the adiabaticity of the collision.
Under some circumstances it appears that one should add (4) a {\it dispersal phase},
since the original clouds may be largely converted to WIM (Fig. \ref{fcat0}d).

In the absence of self-gravity and for Mach number high enough
that Mach scaling applies (KMW) the primary 
parameter controlling head-on collisions between identical clouds is 
\begin{equation}
\eta=\frac{N_{rad}}{n_cR_c},
\end{equation}
where $N_{rad}$ is the radiative cooling column density, while $n_c$ and $R_c$ 
are the cloud number density and radius, respectively. Note that this definition
for $\eta$ agrees with the one in eq. (5) of RFM; their adjustable parameter
$\alpha$ is here taken equal to unity. Thus when comparing the two sets of results
our $\eta$ should be divided by $\alpha^{-1}\sim 3$.
If we express $N_{rad}$ 
in terms of the radiative cooling time $\tau _{rad}$, $n_c$ and the individual cloud velocity
$v_c$ ($N_{rad}=n_c v_c \tau _{rad}$),
\begin{equation}
\eta=\frac{\tau_{rad}}{R_c/v_c}=\frac{\tau_{rad}}{\tau_{coll}},
\label{eta2}
\end{equation}
where, according to Spitzer (1978), 
\begin{equation}
\tau_{rad} = \frac{3}{2}\frac{kT}{n\Lambda(n,T,Z)}.
\label{trad}
\end{equation}
and $\Lambda(n,T,Z)$ is the interstellar cooling function depending on 
number density $n$, temperature $T$, and metallicity $Z$.
We have introduced the collision time $\tau_{coll}=R_c/v_c$, which
is a natural timescale for \cc~and is about the time over which the 
compression phase occurs. The collision is adiabatic if 
$\eta \gg 1$, radiative if $\eta \sim 1$ and isothermal if
$\eta \ll 1$. 
Since according to eq. \ref{trad} $\tau_{rad}$ only depends on the
density and temperature of the cloud (and on the metallicity, $Z$),
then, from eq. \ref{eta2} $\eta\propto v_c/R_c$ and we can infer that small,
supersonic clouds undergo primarily adiabatic collisions,
whereas interactions between large, slow clouds are
mostly radiative or isothermal.
Since in the adiabatic case \coll$\ll\tau _{rad}$,
there is not enough time for the thermal energy to be radiated away during 
the collision. 
On the other hand, in the radiative regime $\tau _{rad}\sim\tau_{coll}$,
and during the collision a significant fraction of the energy associated with
the clouds is converted into radiation. This fraction is even larger
in the isothermal case.

Occasionally in calculations such as the present ones, strong radiative
cooling is taken into account approximately by setting the gas adiabatic
index $\gamma\sim 1$. That is, the flow is assumed to be
inherently isothermal.
This method allows much greater
compression in the flow than the usual $\gamma =5/3$ case, an effect similar to
that expected during a radiative compression. 
However, this approach {\it is not appropriate}
for the problem at hand. As long as total energy is conserved, even a
$\gamma\sim 1$ gas will have a substantially higher pressure behind a
strong shock than in front of it. That leads in the present
situation to strong pressure
gradients that drive gas flows out of the impact region during a collision. 
On the other hand,
when radiative cooling is included properly, the thermal energy 
is removed  from the interaction region (during the collision) before the 
re-expansion takes place, significantly reducing forces that
drive gas away from that region.
We carried out comparison simulations using  $\gamma\sim 1$ to
approximate strong radiative cooling. They had properties
very different from the properly radiatively cooled flows, actually
resembling more nearly the adiabatic flows for the reasons already
mentioned.

Considering weakly radiative or fully adiabatic cases, on the other hand,
it is worthwhile to notice that
we performed several tests with $1.5\le\eta\le\infty$ and did not
find any substantial differences among them either in qualitative
or quantitative terms. This means that even though our adiabatic 
simulations are characterized by $\eta \gg 1$,
they also represent reasonably well cases with $\eta \ge 1.5$.

In this paper we restrict our study to head-on collision of
neutral hydrogen (HI) clouds, and only consider supersonic clouds.
In fact, according to Spitzer (1978), supersonic clouds should be the most 
common case for the ISM.
We neglect for now the magnetic field. 
KMW and RFM included, in their calculations, a magnetic pressure term
corresponding to an initially weak ($B\sim 1\mu G$) magnetic field 
in order to limit the extraordinary compression otherwise occurring 
during radiative or isothermal collisions. However, a full 
MHD simulation was lacking. In complex flows, shear is typically
more important to magnetic field behavior than compression 
(\eg Frank \etal 1996, Jones \etal 1996), so full MHD may be expected
to behave differently.
We ignore any thermal conduction effects, as well as self-gravity
in the present simulations.

We chose three different conditions for the clouds at impact, and for each investigated 
adiabatic and radiative flows, giving a total set of six cases.
The simplest initial condition involves uniform, pressure-bound
clouds, immediately adjacent and placed in motion at the start of the 
simulation in an initially uniform background. Thus, these collisions take
place before the clouds have formed any structure due to their
motions; that is, they are ``non-evolved''. Although this case is not
very realistic, it most closely resembles previous work (\eg, Stone 1970a, 1970b,
LMPS, KMW)
and, since it produces clean demonstrations of the four stages
of \cc, it is very useful as a ``standard model''.
Next, to add some realism and to include nonuniform structures in a natural
way, we also considered collisions between clouds that had evolved
independently before impact. Those clouds were, otherwise, identical
to the ones used in the ``standard'', ``non-evolved'' collisions.
We considered two varieties of such ``evolved'' clouds. First we
followed collisions between two clouds after each had undergone an
identical evolution time (being a so-called ``crushing time'', defined
in equation \ref{crush}), so that the impacting clouds still had a
mirror symmetry. 
We alternatively allowed collisions between two clouds of somewhat  
different evolutionary ages. Since during their evolution the
clouds become increasingly irregular due to KHIs and RTIs, these
last collisions involve ``non-symmetric'' clouds. Thus, we are able
in a simple, but natural way to begin exploring the r\^ole of
asymmetry in collisions. Properties of the clouds used in each
simulation are summarized in Table \ref{tbl-1}.

The natural timescale for individual supersonic cloud evolution is the so-called
``crushing time'', $\tau_{cr}$,
(\eg KMW, Jones \etal 1994) defined as
\begin{equation}
\tau_{cr}=\frac{2R_c\chi^{1/2}}{v_c} = \frac{2R_c\chi^{1/2}}{Mc_{si}}=1.3
\left(\frac{R_c}{pc}\right)
\left(\frac{\chi}{100}\right)^{1/2}
\left(\frac{M}{1.5}\right)^{-1}
\left(\frac{c_{si}}{10km\,s^{-1}}\right)^{-1}~Myr.
\label{crush}
\end{equation}
During an interval $\tau_{cr}$ a cloud moving through an external medium
develops a bow shock and is
maximally compressed by an internal shock originating at the front part.
An extended, low-pressure wake develops behind the cloud.
Also on this timescale, KHIs and
RTIs start to disrupt the
cloud as it begins to become decelerated with respect to its
background (\eg Vietri \etal 1997). 
In the absence of magnetic fields, simulations show that clouds disrupt
because of instabilities on timescales $t \ge \tau_{cr}$.
Detailed discussions of the physics of individual, supersonic  cloud
evolution may be found in Doroshkevich \& Zel'dovich (1981), Jones \etal (1994),
Schiano \etal (1995), Jones \etal (1996), Vietri \etal (1997).
For the clouds considered here, $\tau_{cr}\sim 5.3\times 10^5yr$
 in the adiabatic case ($R_c=0.4pc$) and 
$\tau_{cr} \sim 2\times 10^6 y$ in the radiative case ($R_c=1.5pc$).
These are of
the same order as the mean time for a cloud to have a collision in the ISM
(Stone 1970, Spitzer 1978, Hausman 1981). This further supports our 
feeling that cloud evolution prior to the collision must be considered.
In our study we begin the individual cloud evolution at  
$t = -\tau_{cr}$ for ``symmetric, evolved'' \cc.
For ``non-symmetric, evolved'' \cc~
one of the clouds begins its evolution at $t = -1.5\tau_{cr}$ before
the collision event. For these purposes, we define the collision
to begin (\ie $t = 0$) when the bow shocks of the two clouds osculate.

\section{Numerical Setup}

\subsection{The Code}

We simulate the CC~problem 
using a 2D Eulerian hydrodynamical code on a Cartesian grid.
The code we used is based on an explicit ``TVD'',
conservative finite
difference scheme, second order accurate both in time and space
(Harten 1983, Ryu \etal 1993).
Multidimensional flows are handled
by the Strang-type dimensional splitting (Strang 1968). 
We accounted for radiative cooling  in each time step 
by explicitly correcting the total energy after updating  hydrodynamical
quantities.

Radiative losses are described generally by the following equation:
\begin{equation}
\label{eqen1}
\frac{de}{dt}=-{\cal L}
\end{equation}
where $e$ is the internal energy and ${\cal L}=n^2\Lambda(n,T)-n\Gamma$
accounts for the net loss to radiation against nonadiabatic heating.
The cooling term ${\cal L}$ defines the cooling timescale
\begin{equation}
\tau_{rad} = \frac{e}{{\cal L}}.
\end{equation}
When radiative losses are very high, as they can be during the 
compression phase, the
cooling timescale, $\tau_{rad}$, is comparable to or less than the dynamical time scale which,
by the Courant condition, ordinarily determines the computational time step.
In this case the cooling term is labeled ``stiff''. There are several ways
to handle stiff cooling terms (see, \eg LeVeque 1997). One way is to 
choose the shorter of $\tau_{rad}$ and the time step imposed by the Courant
condition as the computational time step.
However, in some situations this choice may lead to uncomfortably
short computational time steps. On the other hand we could employ 
Strang's operator splitting approach, where the
cooling is computed by multiple steps with its own time step, during one
computational time step determined by the Courant condition.
We chose a third approach in which eq. \ref{eqen1} is
rewritten as 
\begin{equation}
\frac{dln(e)}{dt}=-\frac{{\cal L}}{e} = - \frac{1}{\tau_{rad}}.
\end{equation}
This leads to the solution between time steps $j$ and $j+1$,
\begin{equation} e^{j+1}=e^{j}\cdot \exp\left(-\frac{{\cal L} \Delta t}{e^{j+1/2}}\right).
\end{equation}
For this method the cooling is computed in a single step
even though the cooling time scale is smaller than the dynamical time
scale.
The method ensures that the internal energy is always positive with the
computational time step, $\Delta t$, determined by the Courant condition. 
If one uses the initial value $e^j$ for $e^{j+1/2}$ the radiative 
correction is, however,  only first order accurate in time.
The radiative cooling function we used, $\Lambda(n,T)$, includes 
free-free emission, recombination lines 
as well as collisional excitation lines with a standard
solar metallicity (Z= Z$_{\odot}$),
whereas the heating ($\Gamma$) is provided through ionization and 
photoionization processes and by cosmic rays 
(for a full description see Ferrara \& Field 1994).
We have neglected the effects due to dust grains considered by RFM; this
accounts for the slight difference in the ISM phase properties of our model 
with respect to theirs.

To allow tracking of material that was initially identified with
each of the clouds we introduced a
passive tracer, $S$, usually referred to as ``color'' (Xu \& Stone,
1995). This quantity (actually, one for each cloud) is initially set to 
unity inside each cloud and zero elsewhere.
It represents the fraction of material  inside each cell that
was originally part of one of the clouds.
The evolution of the color is followed with van Leer's second-order
advection scheme (van Leer 1976).

\subsection{Grid, Boundary Conditions and Tests}

In each simulation the clouds are centered on the x-axis with reflection
symmetry assumed across this axis.  Only the plane $y\ge 0$ is included
in the computational box.
Tests with this code show that this more economical grid gives
results equivalent to those obtained with a full plane.
The length scale is adjusted for each case so that $R_c = 1.0$.
In those units, the computational domain is $x$=[0,20] and $y$=[0,10].
The grid is Cartesian, so that our clouds are actually cylinders.
The left, top and right boundaries are open. Reflections 
at these open boundaries are too small and too far away from the
collision to affect the structure of the flow.
Test calculations with a computational box twice as
large showed no relevant differences from the results we describe below.
We have explored a range of numerical resolutions, although
only the computations at the highest resolution are presented here.
These involve a $1024\times 512$ grid,
which provides a resolution of 50 zones across the initial cloud radius.
We have also compared our results on a uniform $512\times 256$ grid
with the Adaptive Mesh Refinement results of KMW.
Our simulations are purely hydrodynamical, while they added, for
convenience a magnetic pressure (but no other magnetic effects) to
limit compression behind radiative shocks.
Still, for the same parameters, we obtain results consistent with theirs
in terms of main hydrodynamical feature
development. 

\subsection{Numerical Values, Physical Parameters and Initial Conditions}

The parameters involved in the problem are quite numerous and
a comprehensive study is beyond our present scope.
However, since we are interested in assessing the importance 
of radiative losses, pre-evolution of clouds through the intercloud medium and
symmetry  of the problem, we have decided to restrict as much as possible 
the parameters' space by developing an accurate model for the multi-phase structure
of the ISM, in agreement with observational data. We then adopt the 
most typical values for cloud and WIM physical properties as derived from such a 
model, hoping that they are truly representative of the ISM conditions. In the
following we give an outline of model assumption. 

As mentioned in \S 2 we consider clouds that are initially uniform and in
pressure equilibrium with the inter-cloud medium.
We set the initial density contrast $\chi=n_c/n_i$=100, where $n_c$ and
$n_i$ are the cloud and inter-cloud number densities respectively, and $n_c  =
22 cm^{-3}$ ($\Rightarrow n_i = 0.22 cm^{-3}$); the cloud temperature
is $T_c = 74 K$  ($\Rightarrow T_i = 7400K$).
This particular choice is dictated by the radiative cooling function adopted in our
calculation, by pressure equilibrium assumption, and by 
the density contrast between the two different phases. 
The equilibrium thermal pressure for the ISM turns out to be
$p_{eq}/k_{\tiny B}=1628 K~cm^{-3}$. 
Each cloud has an initial Mach number $M=v_c/c_{si}=1.5$,
where $c_{si}$ is the sound speed in the inter-cloud medium.
The adiabatic index $\gamma=5/3$ ($p = [\gamma-1]e$)
is assumed throughout the calculations. 
For the adiabatic cases we set $R_c = 0.4 pc$, whereas
for the radiative ones, $R_c = 1.5 pc$.
With this choice of the parameters we have $c_{si}\approx 10 km~s^{-1}$~
and $v_c\approx 15 km~s^{-1}$; $\tau _{coll}\approx 2.6\times 10^4 yr$
for the adiabatic cases. 
For the radiative cases $\tau_{coll} \approx 9.7\times 10^4 yr$ 
and the radiative cooling time inside the clouds turns out to be
\rad $\approx 3.7\times 10^4~yr$, yielding $\eta\approx 0.38$.
Finally, the Jeans length associated with the initial clouds is 
$\lambda_j\approx 29 pc\gg R_c$. In the radiative symmetric collisions,
the large density increase produced during the compression phase,
causes a significant reduction of $\lambda_j$, which 
becomes comparable to, yet still larger than, the vertical size of 
the clouds. For this reason we have neglected 
self-gravity throughout our calculations (see also KMW).
However in a more refined calculation, which would take into account 
other processes like chemical reactions or recombination processes,
larger compression might be allowed making, as a result, the colliding clouds
gravitationally unstable.

As explained in \S 2 we have allowed the described adiabatic and
radiative clouds to collide 
under three different circumstances. For Cases 1 and 2 in Table \ref{tbl-1},
uniform clouds are placed on the grid in such a way that their initial
boundaries are only 2 zones apart at $t = 0.0$. 
These are the so-called ``non-evolved''
cases. For Cases 3 and 4 each cloud begins on independent evolution at
$t = - \tau_{cr}$ (as appropriately determined by their 
properties listed in Table \ref{tbl-1}). 
In these cases $t=0.0$ is defined as the moment when the bow shocks of the
clouds osculate.
For Cases 5 and 6 one of the clouds begins its independent
evolution earlier, at
$t = -1.5 \tau_{cr}$. Again, the two cloud bow shocks 
come together at $t = 0.0$. Table \ref{tbl-1}
lists these details. Animations of each simulation have been posted on the 
World Wide Web site at the University of Minnesota.

\section{Results}
\label{resu}

\subsection{Collision of non-evolved clouds}
\label{cat0}

Fig. \ref{fcat0} shows the four  phases defined in \S 2 for an adiabatic
collision (Case 1).
At the earliest time shown (Fig. 1a), \tt = 1.5\coll, the
collision is near the end of the compression phase.
At the very beginning of this phase
a one-dimensional analysis in the limit of strong shock can still be
applied to the shocks propagating through the clouds.
Theory predicts $\rho\sim 4\rho_c$ and 
$p\sim (4/3)\rho_c v_c^2$, in very good agreement with the numerical values
found (Fig. \ref{cut}). 
The high pressure in the interaction region
limits the compression and leads to a fast re-expansion.
The compression phase lasts
longer near the cloud centers, because the cloud column density is maximum 
along $y=0$, and
because some gas is vertically squirted out from side edges of the interaction
region, right after the
beginning of the collision (Fig. \ref{fcat0}a). The ejected
material propagates through the lower density WIM,
and later on develops
features commonly seen in astrophysical jets. In particular
this slab-jet structure shows a working surface bounded by a shock,
a cocoon surrounding the jet and apparent KHIs.
Nevertheless, since this gas represents
a small fraction of the total mass of the clouds, it does not affect the
development of the collision very much. At the end of this phase
the gas is highly compressed into a layer much
thinner than the initial cloud size (Fig. \ref{fcat0}a and \ref{cut}).

After \tt = 1.5\coll, as already pointed out in \S \ref{descr},
the shocks generated within the clouds during the collision enter the WIM and 
allow the clouds to re-expand. 
Re-expansion takes place 
supersonically, generating a shock 
that, with the jet-shock, develops a nearly circularly expanding
shock-structure on the x-y plane. 
Inside this structure a reverse shock is generated and the
re-expanding cloud material begins to accumulate in a dense 
shell with $\rho\sim 30\rho_i$ (Fig. \ref{cut}).
By \tt = 8.2\coll~ 
(Fig. \ref{fcat0}b)
a dense layer is
well-formed and, despite its expansion, has become the region of highest 
density ($\rho \sim 10-
16\rho_i$). It has a nearly circular shape, except for distortions by 
KHIs and RTIs, which eventually form long dense
($\rho\sim 11.5\rho_i$) fingers. The shell
expands at pretty high velocity ($v\sim 1.1 c_{si}$). 
After about \tt=12 \coll~ the re-expansion of the shell halts, just
before the reverse shock passes from the shell into the central,
low-density cavity.
Then the collapse of the shell begins. 
Fig. \ref{fcat0}c shows the collapse phase at \tt = 37.5\coll.
Eventually, the reverse
shock reflects off the x-axis and rebounds. But, it is then largely
disrupted by refraction in the irregular density structure of the
collapsing cloud material. Large-scale vortices are generated that
through KHIs and RTIs hasten the formation of complex filaments evident 
by \tt$\sim$ 67.5 \coll~ (Fig. \ref{fcat0}d).
At the end of the simulation
what remains of the two clouds is a very low density central region with
$\rho$ mainly between 2$\rho_i$ and 3.5$\rho_i$ (Fig. \ref{cut}), surrounded by 
a complex of filaments, which are mixing the original cloud material with the
WIM. {\it We deduce that, in 
Case 1, the likely fate of the clouds is disruption and conversion of
cloud material into the WIM phase.}

The analogous radiative Case 2 is shown in Fig. \ref{fcct0}.
During the compression phase (Fig. \ref{fcct0}a) a fair
fraction of the thermal energy is radiated away (Fig. \ref{plot-rad}).
As a consequence, the
density reaches very high values ($\rho\sim 10^4\rho_i$) and 
re-expansion is much slower than in Case 1.
Since the re-expansion is so slow now,
the reverse shock promptly
penetrates all the way back to the impact surface and is reflected outwards
again. This sequence, much like what happens inside a young supernova remnant
(\eg Dohm-Palmer \& Jones 1996),
eliminates the central low pressure region.
Following this, a significant re-expansion along the initial direction
of motion gives back to the merged cloud material a typical
cloud-like aspect ratio. The structure undergoes some vertical expansion too,
but the most prominent feature in this direction is a thin jet propagating
along the symmetry plane of the collision.
This is shown in Fig. \ref{fcct0}.
At \tt = 9\coll~ (Fig. \ref{fcct0}b), almost all the 
cloud gas is still in a core with high density ($\rho\sim 580\rho_i$),
surrounded by lower density material with $\rho$ ranging between 200 
and 450$\rho_i$,
and expanding at a very low speed. A comparison of Fig. \ref{fcct0}b with 
Fig. \ref{fcat0}b shows clearly that the size of the re-expanded cloud is
much larger in the adiabatic case than in the radiative one. As a consequence,
in the radiative case the slab-jet structure is much more distinct.
For Case 2 the simulation ends at \tt = 37.5\coll. {\it We deduce that in Case
2 the likely fate of clouds is coalescence.}
The apparently coalesced clouds have evolved into an almost circular 
object of radius $\sim 2 R_c$ with densities ranging between about
20$\rho_i$ in the inner part and 70$\rho_i$ at the surface.
Its total mass is $M_{tot}\simeq 0.84\times 2\times M_c$, where $M_c$ is the
initial cloud mass, showing a high efficiency (84\%)
for the buildup mass mechanism.
The edge of the newly formed object is sharply bounded, but shows
clear signs of KHIs. However, because of the high density of its 
external layer, KHIs will become effective on a timescale much
longer than in the adiabatic case.
The expansion velocity inside the merged cloud at the end is small;
namely a few $\times10^{-2}c_{si}$. 
The net radiative cooling is positive (which means on balance that the gas
is losing thermal energy) in the outer, denser part
of the merged clouds, and negative (which means that the gas is being heated up by
the background radiation and cosmic rays) in the inner more diffuse region, although
in both cases the energy gains or losses are not very significant.
There is a small outward-facing pressure gradient within the cloud
concentration, so eventual collapse seems likely.
Although coalescence seems likely in the near-term,
the final fate of the cloud is unclear.
KMW suggest that the cloud will expand and contract
multiple times
until pressure and thermal equilibrium are reached,
developing filamentary structures along the x-axis in the process.
It might also be possible that the lower pressure inside the cloud
induces contraction followed by a sufficient increase in the
density to turn the cooling function positive. In that case
the extra pressure due to in-fall could be radiated away and the original
cloud density, for which radiative equilibrium holds, might be approached.
In addition, because of the highly dense external layer around the newly
formed structure, we do not expect in the radiative case that RTIs
will be as disruptive during the collapse phase as for the
adiabatic case.

\subsection{Collision of evolved clouds: symmetric cases}
\label{caid}

Fig. \ref{fcaid}a shows the initial conditions for the adiabatic collision
of two evolved clouds (Case 3). At $t = 0$, when the cloud bow shocks
just touch and each cloud has evolved through one crushing time,
the density changes smoothly through the clouds, 
ranging from 10$\rho_i$, at the back, to 150 $\rho_i$ at the front part.
The x-component of the velocity follows the same pattern, being higher
at the front of the cloud than at the rear, although the range of this variable 
is much smaller (\eg Jones \etal 1994). 
On average the clouds have an individual speed corresponding to a Mach
number, $M = 1.35$, relative to the WIM.
There are a number of differences introduced into the interaction
between the clouds by allowing for prior evolution. The most important
are the presence of bow shocks and incoming
gas motions within each cloud wake once the clouds collide.
After the approaching clouds encounter each other's
bow shock, reverse shocks (which act as secondary bow shocks) are generated.
In the adiabatic interaction (Case 3) these shocks substantially affect the clouds.
In fact at \tt = 2.2\coll, right before the cloud bodies encounter the bow shocks,
their x-width $\ell\sim 1.5 R_c$ and $\rho_{max}\sim 150$, but at \tt = 3\coll, 
after the bow shock-cloud collision begins, 
$\rho \sim 250\rho_i$ (\rmm$\sim260\rho_i$) and $\ell\sim 1.2 R_c$. 
There is further compression so that
at \tt$\sim$ 3.8\coll,
right before the cloud bodies impact each other,
$\rho \sim 380\rho_i$ in the compressed front layer, and $\ell\sim R_c$ (Fig.
\ref{fcaid}b). 
So, the maximum compression reached during the collision (\tt$\sim$ 4.5\coll)
(\rmm$\sim 1200\rho_i$) is much higher in Case 3 than in Case 1,
although the pressure enhancement is about the same, and both
collisions are adiabatic.
If the clouds were self-gravitating, such differences might become
important to the possibility for triggering star formation out of
such collisions.
On the other hand, the jet-like and the thin shell structures develop in
roughly the same pattern as for Case 1.
By comparison to the non-evolved collision, however, more cloud material
remains in a core structure, almost to the re-expansion phase.
This is evident in Fig. \ref{fcat0}b,
at \tt = 11.2\coll, 
when contrasted with Fig. \ref{fcaid}c.
The same comparison also shows that the shell structure in Case 3 is more
irregular than in Case 1. This is in fact due to the lower density of the
shell of Case 3, which allows a quicker development of RTIs and KHIs.
By the time shown in Fig. \ref{fcat0}c expansion along the x-axis
has been reduced compared to the y-direction.
That results from the interaction between the expanding cloud material
and inflowing WIM within the wakes of the two clouds. The ram pressure
of the wake flows 
is strong enough to affect significantly the expansion along this path.
Indeed at \tt$\sim$15\coll~a standing, reverse shock is well formed and 
deflects the expanding cloud material away from the x-axis.
This in turn becomes strongly sheared and filamentary due to KHIs.
A substantial fraction of the cloud material is 
ejected so that it cannot join the collapse.
The shock structures in this case are very complex, since they
involve interactions with the pre-existing bow shocks and tail
shocks and a generally more complex density and pressure structure as
the collision begins.
Nevertheless, as our simulation of Case 3 ends (\tt = 22.5\coll),
the dominant, irregular shell structure resembles qualitatively that in Case 1.
Subsequent evolution in Case 3 should follow a pattern
similar to that in the analogous non-evolved Case 1.
In particular we
expect disruption of the clouds in both cases. 

In the radiative evolved case (Case 4) the individual clouds
show a similar qualitative structure as in Case 3, but now their compression due
to initial motion through the inter-cloud medium is much higher (\rmm$\sim1400\rho_i$).
The additional compression results from enhanced radiative cooling induced by
the shock compression within the cloud.
We note that the bow-shock compression phase that was important to
Case 3 turns out not to produce a very significant
effect for Case 4.
That is because the compression brought on by radiative losses 
prior to the encounter is already very large.
In addition, since the speed at which the bow shock penetrates
each cloud scales inversely with the square root of density, 
in this case the
bow shocks barely penetrate into the clouds bodies before they collide,
producing only little pre-compression with respect to the adiabatic case.
As a result, the compression reached during the collision 
is only slightly higher (\rmm$\sim 4\times 10^4\rho_i$) than in the non-evolved
case (Case 2).
The re-expansion phase in Case 4 follows pretty much the same pattern
as in Case 2, except that now it is substantially slowed down in some
directions by the
action of the wakes behind each cloud, as we noted also for
Case 3. In fact, as shown in Fig. \ref{fccid},
by the end of this simulation (\tt = 22.5\coll) the clouds have merged into a dense structure of size
$\sim 2R_c$, which is 2/3 of the size of the merged core in Case 2. 
Expansion in the y-direction is relatively free and leads to a KH 
unstable jet as in the other symmetric collisions. However, re-expansion
along the x-axis is strongly inhibited by inflowing wake material, so that
the velocity of expansion along the x-axis is only about 1/3 as large, compared
to Case 1.
As a result the expanding gas collects in a high density cloud ``rim''
($\rho\sim 400\rho_i$, \rmm$\sim  440\rho_i$).
The wake material forms a standing, outward facing, 
``accretion shock'' outside the cloud structure. Examination
of the color variable shows a very good correspondence between the high
density material visible in Fig. \ref{fccid} and cloud material.
Thus, the figure shows that wake material, once it impinges on the cloud, joins
the outflow of cloud material in the y-direction.
In fact, it appears that the inflowing wake
material  and associated shocks are responsible for driving
the vertical outflow of cloud material along the jet and for
producing the KHIs that have generated the large eddies evident in Fig. \ref{fccid}.
In the inner cloud the density ranges between 100 and 200 $\rho_i$,
which is higher by almost a factor 2 with respect to the non-evolved case.
Even though there is a weak pressure gradient pushing vertically
in the central condensation, it seems very likely that in an extension 
of this simulation the main core of the merged clouds would remain intact. 
{\it So, we judge these clouds to be coalesced.}

\subsection{Collision of evolved clouds: asymmetric cases}
\label{cadt}

As mentioned earlier we chose a simple, but natural way to explore
symmetry breaking in the collisions just discussed.
That is, we collided clouds that were identical when set into motion, 
but which were differently evolved at impact. On a timescale $\sim$\tcr~the
compression substantially deforms the clouds, while KHIs and RTIs
will produce irregular cloud boundaries.
Since those features are highly time dependent, two clouds of
even slightly different dynamical ages will lack mirror symmetry.
In our asymmetric simulations one cloud (C1 hereafter, and on the
left) was aged by 1\tcr~and the other (C2 hereafter, and on the right)
by 1.5\tcr~as their bow shocks came in contact at $t = 0.0$.
The older cloud had in general a denser front part
and a smaller velocity  at impact (Jones \etal 1994).
However the aspect ratio (length to height ratio) 
developed by the cloud during its motion
through the WIM is strongly related to the adiabaticity of the 
gas. Indeed it increases in the adiabatic case, but 
decreases in the radiative one, so that the radiative cloud grows denser and more compact
as it evolves (Vietri \etal 1997).
This turns out to have a major impact on the survival of clouds in asymmetric
\cc.
As in the previous evolved cases, the two clouds 
undergo bow shock-compression before colliding bodily. 
In the asymmetric interactions the clouds
have different speeds and shapes, and
are located at different distances from their bow shocks.
As a result they no longer experience compression simultaneously.
In addition, since C2's bow shock is weaker than C1's,
bow-shock compression for C2 turns out to be stronger than for C1. 
In the adiabatic Case 5, right
before the direct collision, C1 has $\rho\sim 300\rho_i$ 
and $\ell\sim R_c$ and C2 has $\rho\sim 370\rho_i$ and
$\ell\sim 0.6 R_c$. 
The compression phase is shorter than in Cases 1~-~4, and not all of the
kinetic energy is converted into thermal energy (Fig. \ref{plot-ad}). 
After the collision, the younger, more compact cloud C1, maintains
its identity longer than C2. As shown in Fig. \ref{fcadt}, at \tt
= 8.2\coll~ C1 still has a dense core with $\rho\sim 70\rho_i$ surrounded by a
layer with $\rho\sim 40\rho_i$. By contrast, C2 is being stretched and torn apart
and, soon after, is mostly converted into the WIM.
Nevertheless C1 is undergoing
rapid re-expansion too and, at \tt = 13.5\coll, the density is lower than 15
$\rho_i$ everywhere. Although it is much more irregular than Cases 1 or 3,
in Case 5 we can still recognize a clear pattern in the evolution
of the re-expansion phase, with the formation of an expanding shock wave along with
a reverse shock and a low density shell, heavily affected by KHIs and RTIs. 
So, at the end of the second, re-expansion phase, part of the cloud gas has already 
been converted into WIM, whereas the remaining part is forming an irregular
filamentary structure with $\rho$ ranging between 5 and 10$\rho_i$.
As in the symmetric-evolved collisions, the interaction between
re-expanding cloud material and inflowing wake gas significantly
influences the expansion. As for Case 3 the wake 
confines expansion along the x-axis through a standing, inward-facing shock. 
Also, as in Case 3, the expanding cloud material is deflected 
around the wake in a strongly sheared manner. The broken
symmetry in Case 5 allows C2 material to expand much more
easily, however, because its greater y-extent at impact effectively
``launches'' it over the wake of C1. In addition, since C2 has a
lower column density just above the x-axis, it recoils in
response to the collision. Thus, a substantial fraction of C2
passes over its own wake after the collision.
Those effects add
considerably to the disruption of cloud material into filaments
and probably hastens its dispersal into the WIM.
The resulting flow is very complex and highly vortical. So,
we expect the collapse phase to be very ineffective at collecting
together material into new clouds. {\it Thus, this case seems clearly
much more disruptive that any of the previous ones.}

On the other hand, the radiative Case 6
is characterized by completely new features.
At \tt = 2.2\coll~ 
the region between the two shocks formed by the collision,
appears to be strongly distorted, resembling the structure that develops
during the ``bending mode instability''
(KMW, Vishniac 1994, Hunter \etal 1986).
As shown in Fig. \ref{fccdt}a, at \tt = 3\coll~ the layer between the two shocks
has grown more corrugated and the front parts of the clouds are following
the same pattern.
Although the bending mode is unstable, the disruption of the 
impact surface is primarily due to the gross irregularities of 
clouds' shape, which dominate the determination of the
evolution of the structure of the flow.
Fig. \ref{fccdt}a shows also that the upper part of the left (C1) cloud
is about to 
break off and pass over the other cloud (C2), carrying some
of C2's gas with it. As already pointed out, unlike the adiabatic case,
in the radiative case the older cloud C2 has a more compact structure,
which makes it more solid and more resistant to the collision than C1.
At \tt = 9\coll~ (Fig. \ref{fccdt}b), 
the material from C1 passing over the top of C2 has expanded
again into a distinct cloudlet, with density ranging from
$\rho\sim 100\rho_i$ at the front, to 
$\rho\sim 5\rho_i$ at the tail, and with still slightly
supersonic velocity ($v_x\ge c_{si}$).
Subsequently, that cloudlet becomes strongly decelerated and suffers
KHI- and RTI- induced destruction, as seen previously for individual
clouds (\eg Jones \etal 1994).
Although the cloudlet leaves our grid before it is destroyed, 
it seems fairly clear that it will dissolve into the WIM.
The wake of the cloudlet is still visible on the far right side in
Fig. \ref{fccdt}c.
On the other hand the core of C2, which by $t = 3$\coll~ has a very long tail and a dense,
but distorted front, passes
through the remaining part of C1, emerging after $t \approx 7.5$\coll~
with the main body of C2 accreted.
This outcome resembles those one would expect in collisions of two clouds of
quite different sizes and densities. It shows that 
the smaller and denser cloud is able to pass through the larger and
more diffuse one, sweeping its gas and
finally breaking it up into two major pieces (Gilden 1984, Kimura \& Tosa 1996).
In these calculations we show that the same fate can actually occur
when two clouds with approximately the same initial characteristics
but with slightly different morphologies collide. 

As in the other collisions between clouds that are followed by wakes,
an ``accretion'' shock forms to the right of C2. However, it is driven
off the grid to the right before the simulation ends. The interaction
is dominated in this case by the high concentration of material in C1.

The merged remnant of C1 and C2 that was formed by \tt=7.5\coll~
has by the end of the simulation, \tt = 30\coll,
evolved into non uniform filamentary structures,
characterized by irregular motion (Fig. \ref{fccdt}c).
Very little mixing between the two original clouds has actually
taken place; rather one has passed through the other.
On the other hand, considerable entrainment of WIM gas has taken place 
through the action of eddies generated during the collision.
The higher density features apparent on the left and top 
perimeter of the main cloud visible in Fig. \ref{fccdt}c are, in fact,
the remnant merged cores of the original clouds, whereas the rest of
the main cloud at this late time contains a strong mix of 
entrained material, or has been ejected from the grid.
At this time the main cloud is being slowly stretched along the x-axis
($v_x\sim 0.1 c_{si}$ at the edges), while large eddies on the top
and downward pointing pressure forces are being effective at reducing 
its height. It seems likely that
the single dense region visible in Fig. \ref{fccdt}c will be
bisected into two before too long. 
The final outcome may be two distinct clouds formed largely from material
originally in C2, which was the more compact of the original pair.
Thus the outcome is completely different from that of Cases 2 and 4
and {\it we conclude that the two clouds are destroyed by
the collision and converted into several filamentary structures.}

\section{Discussion}

We have investigated for the six cases summarized in Table \ref{tbl-1}
the collision of diffuse HI clouds in a multi-phase medium.
Our objective is to understand such issues as the likely fate of clouds after 
collisions, including the conditions for coalescence and the fraction
of the initial kinetic energy radiated away. 
In the previous section we outlined the basic dynamical
evolution of each collision and the ultimate 
fate of the clouds. It was clear from those examples that the
fate of colliding clouds depends strongly on the symmetry of the
interaction and also on the degree to which the initial kinetic
energy is radiated away before the clouds begin their re-expansion.
To clarify and expand on those issues we now review the main points,
separately for the adiabatic and radiative cases.

\subsection{Adiabatic Cases}

Adiabatic collisions generally appear to result in cloud
disruption, with most of the gas converted into the WIM phase.
This point is made clearer in Fig. \ref{plot-ad}, which
shows a plot of various properties characterizing individual clouds involved in
adiabatic collisions. In particular, using the color
variables, we can follow the kinetic and
thermal energies of each cloud, normalized to the initial cloud total energy,
as well as the center-of-mass coordinates
$x_{cm}$ and $y_{cm}$ for each cloud relative to the point of first
contact
(see Jones \etal 1996 for a mathematical definition of $x_{cm}$ and $y_{cm}$).
From a comparison of
panels (a) and (b) in Fig. \ref{plot-ad}, we see that in all adiabatic Cases 1, 3 and 5 the kinetic
energy of the clouds $E_{kin}= 1/2M_c v_c^2$ is initially converted  mostly
into thermal energy $E_{ther}= pV/(\gamma - 1)$
during the compression phase, through the action of
the main, outward moving shocks as they pass through the cloud bodies. 
That stage is immediate for the ``nonevolved'' collision, but is
delayed, of course, for the ``evolved'' cases, since \tt = 0 corresponds
to the moment when the bow shocks touch, rather than when the clouds first
touch.
During the re-expansion phase, some of this thermal energy is converted back into
kinetic form, since the rapid expansion of the ``blast wave'' into the
WIM, reduces the pressure around the merged clouds.
But, because the cloud shocks have generated entropy,
the cloud gas has substantially more thermal energy at the end of each simulation
than at the start,
despite considerable volume expansion.
This effect enhances the tendency for the cloud material 
to be converted into the WIM phase following the collision.
Also, these plots show that the normalized kinetic
energy decreases gradually with time; this is due to the irreversible 
work done by the
expanding gas on the surrounding background.
A similar trend was observed in the 3D simulations of two colliding
gas streams by Lee \etal (1996).
It is important to notice
that at late times (\tt$\ge$8\coll) some energy decrease
is also due to escape of matter from the computational domain, particularly
the top. 
The center-of-mass positions provide a quantitative measure of this 
expansion as shown in panels (c) and (d).
In particular, note for Case 1 that the maximum in $y_{cm}$ around \tt$\ge$8\coll~
corresponds to the time when the dense shell reaches the top of
the grid.
After that time $y_{cm}$ represents only mass the remaining within the grid.
Panel (c) also illustrates quantitatively how cloud material in the
symmetric collisions is more efficiently removed from the central
interaction region than in the asymmetric case.
Notice for Case 5 that some cloud material remained in the interaction
region interior.
In panel (d) we also see the clear difference in y-expansion of the
individual clouds in the asymmetric Case 5. The more evolved cloud, C2,
which begins the simulation having a greater height and lower
column density along central impact axis, is obviously
disrupted by this measure. The more compact cloud, C1, remains
compact in this dimension, on the other hand. 

\subsection{Radiative cases}

In the radiative cases,
as long as the initial geometry is symmetric, the two colliding 
clouds merge and it seems likely that the conditions
for a new massive cloud to form do exist. As shown in Fig. \ref{plot-rad},
the kinetic energy of the two 
clouds is converted into thermal energy and soon radiated away. 
In deed, unlike the adiabatic cases,
after the compression phase there is no re-enhancement of the kinetic energy,
as is shown by a comparison of panels
(c) and (d) in Fig. \ref{plot-rad} and \ref{plot-ad}.
As a result the re-expansion of the gas is much reduced with
respect to the adiabatic case.
At the end of the evolution the thermal energy of the clouds has increased,
due to heating processes occurring during
the re-expansion phase (Fig.  \ref{plot-rad}).
Also, in the
evolved (symmetric) cases, the density at the compressed layer formed during 
the collision is higher than in the non-evolved case and this
could be important for triggering star formation processes.
Furthermore, the wakes developed during the cloud evolution
limit the re-expansion, so that after the collision
in Case 2 the merged clouds end up forming a smaller and denser region.
In cases where self-gravity
is dynamically important, the wake effect could play a fundamental 
r\^ole, reducing the expansion velocity of the material below the escape 
threshold, thus 
making it possible for the two clouds to build up a new larger,
gravitationally bound structure (see, also, Dinge (1997) for similar
behavior in a single, moving, self-gravitating cloud).

The outcome of \cc~is very different for the asymmetric, i.e. differently 
evolved clouds, even though our collisions all involve clouds of
equal mass and head-on collisions. 
In this case we obtain fragmentation of the ``younger'' cloud 
and the formation
of a small expanding cloudlet eventually dissolving into the WIM.
As shown in Fig. \ref{plot-rad}, in this case only part of the
initial kinetic energy is radiated away during the
collision. As a result, the remainder of the merged cloud material
re-expands and, through interaction with the wake flow, becomes
concentrated into dense clumps.
At the end most of the cloud material has been converted into WIM,
its directed kinetic energy partly being lost
to radiation, partly to turbulent gas motion.
KMW have shown that by slightly perturbing
the surface of one of the two colliding clouds, the bending mode
instability causes fragmentation rather than coalescence.
In our study the asymmetry 
in the problem is due to a slight difference in the evolutionary 
ages of the clouds (0.5\tcr). As already pointed out, its
effect is comparable to that of an off-center collision.
That is, the mass and momentum of the two clouds at impact have 
significantly different $y$ distributions.
Those structural differences of the clouds
make the outcome of the collision very different from the symmetric cases.
From this fact we conclude that supersonic radiative collisions of clouds
with the same gross characteristics (mass, speed and structure),
but without high symmetry,
are disruptive and generate irregular filamentary clumps.
On the other hand previous results show that collisions of clouds with
major structure differences (density, size), are likely to be
disruptive as well (Hausman 1981, Gilden 1984, Kimura \& Tosa 1996, RFM).
Our results support this.
In addition, our results show that the adiabatic
case, which applies to small clouds, will be highly disruptive, even
if the clouds are identical and the collision head-on.
Also, as pointed out below, off-center collisions, which are the most
common, certainly are very unlikely to produce coalescence of clouds, at 
least for non-self-gravitating objects.

These results have an important impact on the ISM of galaxies. 
In general, supersonic, gasdynamical \cc~tend to be disruptive,
so that any model to explain cloud evolution and mass spectrum 
has to take these findings into account.
If colliding clouds are sufficiently electrically conductive, then the presence of
a large-scale magnetic field may play an important dynamical r\^ole
in the interaction. It is not obvious, however, that our main
conclusion will be altered. We have fully MHD simulations underway
that will address that point in a separate report.

It is worthwhile to mention that our asymmetric calculations
can provide some insights for off-center collisions as well, at least for those
with small impact parameter ($b\ll R_c$). In fact, with regard to 
survival of the clouds, the most important implication  
for small impact parameter off-center collisions is probably asymmetry. 
For these cases the results of our asymmetric calculations may apply as a guide,
at least for clouds of comparable mass.
However, when the impact parameter $b\sim R_c$, only part of the
cloud is involved
in the collision and our calculations are not appropriate anymore.
LMPS have investigated off-center, isothermal collisions for very
massive clouds, including gravity in their calculations.
They show that, for high relative velocity, the colliding parts 
of the clouds soon re-expand and disperse after the compression phase,
whereas the outer parts (which do not get involved in the collision)
proceed unhindered
and form two new small clouds. On the other hand, for low relative velocity,
the colliding clouds coalesce, whereas the outer gas motion is deflected into 
a circular pattern. As a result rotating bound systems and bars form.
For smaller non-self-gravitating clouds, we think that, in addition to the
relative velocity, the adiabaticity is a key parameter. If $b\leq R_c$
the collision can probably be classified in similar terms to a highly 
asymmetric one. 
On the other hand, if $b\ge R_c$, the LMPS calculations suggested 
that the outer part of the clouds are torn apart during the collision.
For an adiabatic collision, pressure waves in the remaining
clouds generated during the encounter might be able to make the clouds 
expand and disperse into the WIM. However, in a strongly radiative 
encounter these waves could be damped away by radiation and the cloud cores 
might survive, although the clouds themselves would turn out quite distorted
Finally, if $b\sim 2R_c$, the collision will produce only minor perturbations
on the cloud structures.

Real clouds are, of course, three dimensional. CC three-dimensional calculations have 
been performed by 
Hausman (1981), LMPS and Lattanzio \& Henriksen
(1988, LH hereinafter); these authors, using a smoothed particle hydrodynamics code,
investigate on the effects of several parameters
(relative velocity, cloud mass ratio, impact parameter and so on) on coalescence.
Hausman's calculations are strongly affected by the limitations of his
computational
means. In particular he used a resolution as low as 100 particles per cloud,
and his calculations show unphysical particle interpenetration. For this reason,
for example, in his run 1, Hausman finds a ratio $p_{max}/p_{ext}$ of only 3.55, much less
than 50 as found by LMPS for the same case. As a result, little conversion
of kinetic energy into thermal energy takes place; it is probably for this reason
that, in most of his runs, Hausman finds that the cooling is efficient enough to ensure
isothermality (Hausman 1981). Since the set of cases studied by Hausman is very similar
to that presented by LMPS, we will neglect to go in more details about his results.
LMPS employed a better version of smoothed particle hydrodynamics code
(Monaghan \& Lattanzio 1985), and higher resolution ($\sim 2000$ particles per cloud).
We have already summarized their results regarding off-center collisions.
They also found that symmetric head-on collisions generate a single rapidly
re-expanding or collapsing cloud depending on whether the initial clouds
are gravitationally stable or unstable respectively. In asymmetric
(but still head-on) cases however, they concluded that even when
a cloud is marginally stable to gravitational collapse, the collision
with a smaller cloud  (ratio of masses larger than
2.5; their clouds had the same density),
does not initiate the instability.
Finally LH further investigated this problem showing how
spin and orbital angular momentum are important in determining the outcome
of \cc.
The accuracy of these results, although useful and to some extent
in agreement with previously cited works and our own,
is still limited by low resolution ($\sim 2000,~3000$ particles)
and  some assumptions which are not always appropriate.
Some of these were mentioned by the authors themselves.
In addition, based on Hausman (1981) results, both LMPS abd LH
assumed isothermal clouds, using $\gamma = 1$;
 but as already pointed out, this
approach is not correct because of the fundamental dynamical r\^ole played
in the re-expansion phase by thermal energy stored during the compression
phase.
That aspect is apparent in our results.
In our radiative head-on symmetric simulations, which allow
for a release of energy through radiative processes, the collisions
produce a new merged stable cloud instead of a rapidly re-expanding cloud
as found by LMPS.
Also LMPS and LH do not include in their calculations an
intercloud medium. However, as we  have noted,
the interaction of cloud material with the intercloud gas, particularly
with shocks and cloud wakes,
significantly affects the evolution of the collision.

Useful insights about how an additional degree of freedom  in this
problem might modify 2D
results may be provided by 3D studies of some related problems.
For example, Xu \& Stone (1995)
examined the evolution in 3D of a gas cloud overrun by a plane
shock. They found behaviors qualitatively consistent with 2D
simulations of that problem (\eg Bedogni \& Woodward 1990; Jones \& Kang 1993;
KMW). In many respects a shocked cloud is similar to
a supersonic cloud in its evolution (Jones \etal 1994).
Lee \etal (1996) have carried out a 3D study of two colliding gas
streams. That is similar in some ways to collisions between
clouds. They found, as we do for adiabatic collisions, that the 
bulk of the kinetic energy is converted into thermal pressure,
and that this causes the colliding material to expand as it
drives a shock into the ambient medium. Generally, the extra
dimension will allow more complex motions to develop, and some
considerations will depend quantitatively on the third
dimension, but we see no evidence in the existing literature that 
the general conclusions of the present work will be invalidated when it 
is included.

\section{Conclusions \& Summary}

To summarize, in this paper we have found the following results for the
gasdynamical collisions between two mildly supersonic interstellar
clouds, $M\ge 1.5$, of comparable mass:
\begin{enumerate}
\item Supersonic \cc~are most often disruptive. In particular adiabatic 
collisions, which involve small clouds ($R_c< 0.4 pc$, for the standard 
WIM parameters assumed in our calculations), 
turn out always  to be disruptive. 
\item For completely symmetric collisions, strong radiative energy losses
can, however, lead to coalescence of the two clouds. In fact emission
of radiation reduces the thermal energy stored during the compression
phase, preventing a vigorous pressure driven re-expansion.
\item Asymmetry in the clouds at impact greatly enhances the tendency
for clouds to be disrupted during the interaction, even when radiative
cooling is strong. This is true even for a very modest asymmetry. 
In the adiabatic, asymmetric case the clouds are almost immediately
dispersed in the WIM. In the radiative case new filamentary structures are
produced out of the initial cloud material.
\item Future numerical work should not neglect the importance of allowing
the clouds to develop
a self-consistent structure, especially bow shocks and wakes, since these
features strongly influence the interaction and add important hydrodynamical
features. 
In particular, bow shock interaction of the
colliding clouds produces higher compression, particularly in the adiabatic case.
On the other hand the wakes behind the clouds reduce the re-expansion 
along the x-axis, increasing the probability for coalescence.
\end{enumerate}
\acknowledgments

FM devotes his efforts in this work, to the memory of his friend,
Leonardo Muzzi, young artist of deep 
perspective, who inspired his way of studying science.
This work by TWJ and FM was supported in part by the NSF through
grants AST-9318959 and INT-9511654 and
by the University of Minnesota Supercomputer Institute.
The work by DR was supported in part by Seoam Scholarship Foundation.
AF acknowledges hospitality of University of Minnesota where this work started.

\clearpage

\begin{deluxetable}{cccccccc}
\footnotesize
\tablecaption{Summary of 2D-HD Cloud Collisions Simulations \label{tbl-1}}
\tablehead{
\colhead{Case \tablenotemark{a}} & 
\colhead{$\eta$\tablenotemark{b}} &
\colhead{$R_c $ } & 
\colhead{\coll=$R_c/v_c $ } & 
\colhead{\tcr=$2 R_c \sqrt{\chi}/v_c $ } & 
\colhead{$M_r$\tablenotemark{c}} &
\colhead{Clouds ages\tablenotemark{d}} & 
\colhead{End Time\tablenotemark{e}} 
} 
\startdata
1 & adiabatic &  0.4 pc & 2.6$\times 10^4y$ & 5.3$\times 10^5y$ & 3   & 0\tcr - 0~~\tcr & 67.5\coll\nl
2 & 0.38 &  1.5  pc  & 9.7$\times 10^4y$ & 2.0$\times 10^6y$ & 3   & 0\tcr - 0~~\tcr & 37.5\coll  \nl
3 & adiabatic &  0.4 pc & 2.6$\times 10^4y$ & 5.3$\times 10^5y$ & 2.7 & 1\tcr - 1~~\tcr & 22.5\coll\nl
4 & 0.38 &  1.5  pc  & 9.7$\times 10^4y$ & 2.0$\times 10^6y$ & 2.8 & 1\tcr - 1~~\tcr & 22.5\coll  \nl
5 & adiabatic &  0.4 pc & 2.6$\times 10^4y$ & 5.3$\times 10^5y$ & 2.4 & 1\tcr - 1.5\tcr & 22.5\coll\nl
6 & 0.38 &  1.5  pc  & 9.7$\times 10^4y$ & 2.0$\times 10^6y$ & 2.7 & 1\tcr - 1.5\tcr & 30.0\coll  \nl
 
\enddata

\tablenotetext{a}{All models have used $\gamma$ = 5/3,
$\chi = \rho_c/\rho_i = 100$, equilibrium pressure 
$p_{eq}/k_B=1628~K~cm^{-3}$. Also, at equilibrium,
we have $T_i= 7400K$ and
$n_i = 0.22~cm^{-3}$ for the inter-cloud medium and 
$T_c= 74K$ and $n_c = 22~ cm^{-3}$ inside the clouds. 
Furthermore, all the computations
were carried out on a rectangular grid with size $N_x = 2 N_y = 1024$
corresponding to a resolution of 50 zones through the cloud radius.
Left, top and right boundaries are free whereas reflection is assumed at the 
bottom.}

\tablenotetext{b}{$\eta = $\rad/\coll}

\tablenotetext{c}{This is the {\it relative} Mach number and is referred
to the inter-cloud sound speed, $c_{si}\simeq 10 km~s^{-1}$.}

\tablenotetext{d}{The left and right hand columns refer to the left and right
 hand clouds, C1 and C2, respectively.} 

\tablenotetext{e}{The end time is expressed in terms of collision time \coll,
and represents the total time from the beginning of the collision.}

\end{deluxetable}

\clearpage

\begin{center}
{\bf FIGURE CAPTIONS}
\end{center}

\figcaption[]{
Inverted grayscale images of $\tanh(\log(\rho))$ for Case 1;
vector display of the velocity field superimposed.
Panel (a) (top) shows the compression phase at \tt=1.5\coll~with outflow at the 
side of the contact discontinuity. 
Panel (b) (bottom) shows
the re-expansion phase at 8.2\coll, with the formation of a dense shell-like 
structure. Panels (c) and (d) correspond respectively to the collapse phase
at \tt=37.5\coll~
and to the dispersal phase at \tt=67.5\coll. 
The dramatic development of KHIs and RTIs is evident.
\label{fcat0}}

\figcaption[]
{Cut through the grid along the x-axis (y=0.3$R_c$) for Case 1. The top and bottom panel show 
log-plot of density and pressure respectively. 
Solid lines refer to \tt=1.5 \coll, dot lines to \tt=8.2\coll, short-dash
lines to \tt=37.5\coll~ and long-dash lines to \tt=67.5\coll.
\label{cut}}

\figcaption[]{Same grayscale as in Fig. \ref{fcat0}, but 
for Case 2. (a) (top) The compression
phase at \tt = 1.5\coll, and the ejection of material along the collision plane
of the crushed clouds. (b) (center) The re-expansion phase at
\tt = 3.7\coll; the (slab) jet-like structure is well-formed and is showing
features common to astrophysical jets.
(c) (bottom) The structure at  \tt = 37.5\coll, when the two clouds have
merged into a structure with a dense rim.
By this time KHI structures are evident along the jet and on the perimeter of the
merged cloud.
\label{fcct0}}

\figcaption[]
{Same grayscale as in Fig. \ref{fcat0}, but now for Case 3. 
(a) Conditions when the bow shocks generated by the clouds osculate, just prior
to impact.
(b) The two clouds a little before the
actual cloud-body collision. The clouds have been compressed by the
bow shocks.
(c) The re-expansion 
phase at \tt = 11.2\coll. The two clouds are still
distinguishable; also a low density layer has formed around them and is
undergoing strong ablation by KHIs.
\label{fcaid}}

\figcaption[]
{Same grayscale as in Fig. \ref{fcat0}, but now for Case 4, with velocity vectors
superimposed. This image shows the final stage 
of the re-expansion at \tt = 22.5\coll. The two clouds have merged
into a new structure with density decreasing from the 
bottom to the top. Delicate density features have been produced by 
KHIs, enhanced by in flowing wake material deflected along the sides of the clouds.
\label{fccid}}

\figcaption[]
{Same grayscale as in Fig. \ref{fcat0}, but now for Case 5, with velocity vectors
superimposed. This image shows the complicated
re-expansion phase at \tt = 8.2\coll. The cloud originating on the left (C1) is still
recognizable, whereas C2, coming from the right, has been strongly distorted 
from its original form.
\label{fcadt}}

\figcaption[]
{Same grayscale as in Fig. \ref{fcat0}, but now for Case 6.
(a) The compression phase
at \tt = 3.7\coll. The contact surface is quite irregular
and asymmetric, foretelling of the disruption to follow.
(b) The structure at \tt = 9\coll~showing the formation
of a cloudlet breaking off from C1.
(c) Density with superimposed velocity vectors,
representing the re-expansion 
phase at \tt = 30\coll. A large irregular structure has formed and 
eddies are shredding the top part.
\label{fccdt}}

\figcaption[]
{The four panels show plots of the kinetic energy (a - top left), thermal energy
(b - bottom left), center 
of mass X-cm (c - top right) and Y-cm coordinates (d - bottom right),
as a function of time, for each cloud involved in the collision.
For Cases 1 and 3 the two colliding clouds are identical to each other, 
so only one is displayed in each case.
Solid lines correspond to Case 1, dotted lines to Case 3, short-dash lines and
dot-long dash respectively to cloud C1 (left) and C2 (right) in Case 5. The delay of the features
of Cases 3 and 5 with respect to the non-evolved Case 1, is due to the 
delay in the former cases between impact of bow shocks and cloud bodies.
\label{plot-ad}}

\figcaption[]
{Same as in Fig. \ref{plot-ad}, but now solid lines refer to Case 2, dot lines to Case 4,
short-dash lines and dot-long dash respectively to cloud C1 (left) and C2
(right) in Case 6.
Taking care to note the significant difference in the Y-axis scale, 
A comparison with panels (a) and (b) of
Fig. \ref{plot-ad} shows that in the radiative 
cases much less kinetic energy is converted into thermal form. In addition,
as shown by the $X_{cm}$ and $Y_{cm}$ panels ((c) and (d) respectively) and
the significant difference in the y-axis scales of the two figures,
re-expansion is much reduced.
\label{plot-rad}}

\end{document}